\def\USEACHEMSO{0} %

\if\USEACHEMSO1
\documentclass[journal=jctcce,manuscript=article,%
]{achemso}
\else
\documentclass[aip,jcp,notitlepage,
tightenlines,
reprint,%
floatfix]{revtex4-2}
\fi

\usepackage{siunitx}[=v2]
\usepackage[utf8]{inputenc}
\usepackage[T1]{fontenc}
\usepackage{stix}
\usepackage[version=3]{mhchem}
\usepackage{hyperref}
\usepackage{numprint}
\usepackage[american]{babel}
\nprounddigits{2}
\npdecimalsign{\ensuremath{.}}  %
\usepackage{units}
\usepackage[normalem]{ulem}

\usepackage{xifthen}  %
\usepackage{mathtools}
\usepackage{booktabs}
\usepackage{dcolumn}%
\usepackage{caption}
\usepackage{subcaption}
\usepackage{layout}
\usepackage{multirow}
\usepackage{derivative}
\usepackage{array}
\newcolumntype{L}[1]{>{\raggedright\let\newline\\\arraybackslash\hspace{0pt}}m{#1}}
\newcolumntype{C}[1]{>{\centering\let\newline\\\arraybackslash\hspace{0pt}}m{#1}}
\newcolumntype{R}[1]{>{\raggedleft\let\newline\\\arraybackslash\hspace{0pt}}m{#1}}
\usepackage{enumitem}
\setenumerate[1]{label=(\arabic*)}

\usepackage{adjustbox}
\usepackage{lipsum}
\usepackage[dvipsnames]{xcolor}
\usepackage{soul}
\definecolor{myred}{rgb}{1, 0.123, 0.404}
\definecolor{mypink}{rgb}{1, 0.2, 0.745}
\definecolor{mycyan}{rgb}{0.090, 0.667, 0.553}
\definecolor{freshgreen}{rgb}{0.051, 0.796, 0.733}
\definecolor{mygreen}{rgb}{0.184, 0.792, 0.0}
\definecolor{myviolet}{rgb}{0.256, 0.207, 1.00}
\definecolor{myorange}{rgb}{0.9, 0.44, 0.0}
\definecolor{mypink2}{rgb}{0.898, 0.0, 0.451}
\definecolor{mypurple}{rgb}{0.659, 0.251, 1.000}

\def\REDLINE{0} %
\if\REDLINE1
\newcommand{\rladd}[1]{\textcolor{red}{#1}}
\newcommand{\rlremove}[1]{\sout{#1}}
\else
\newcommand{\rladd}[1]{#1}
\newcommand{\rlremove}[1]{\unskip}
\fi

\usepackage{xspace}
\usepackage{verbatim}
\let\oldtheequation\theequation
\makeatletter
\def\tagform@#1{\maketag@@@{\ignorespaces#1\unskip\@@italiccorr}}
\renewcommand{\theequation}{(\oldtheequation)}
\makeatother

\if\USEACHEMSO1
\fi

\newcommand{\myunit}[2][]{%
  \ifthenelse{\isempty{#1}}%
    {\ensuremath{#2}}%
    {\ensuremath{{#1}\:{#2}}}%
}

\newcommand{\matrgreek}[1]{\ensuremath{\pmb{#1}}} %

\usepackage{listings}
\if\USEACHEMSO1
\newcommand{\fighalf}{0.5\textwidth}
\else
\newcommand{\fighalf}{1\columnwidth}
\fi
\newcommand{\figfull}{1\textwidth}

\usepackage[
textsize=small,
]
{todonotes}

\newcommand{\matr}[1]{\ensuremath{\mathbf{#1}}}
\newcommand{\braket}[2]{\ensuremath{ \langle #1 | \, #2  \rangle }}

\newcommand{\ket}[1]{\ensuremath{  | {#1} \rangle}}

\newcommand{\matrixe}[3]{\ensuremath{ \langle{#1} | {#2} | {#3} \rangle }}

\newcommand{\mpsg}[2]{\ensuremath{\mathbf {#1}^{[#2]} }}
\newcommand{\mpst}[1]{\ensuremath{\mathbf M^{[#1]} }}

\newcommand{\mpstr}[1]{\ensuremath{\mathbf R^{[#1]} }}
\newcommand{\mpsm}[1]{\ensuremath{\mathbf M^{\sigma_{#1}} }}
\newcommand{\mpsml}[1]{\ensuremath{\mathbf L^{\sigma_{#1}} }}
\newcommand{\mpsmr}[1]{\ensuremath{\mathbf R^{\sigma_{#1}} }}

\newcommand{\lstatei}[2][]{\ensuremath{ l_{\alpha^{#1}_{#2}} }}
\newcommand{\rstatei}[2][]{\ensuremath{ r_{\alpha^{#1}_{#2}} }}

\newcommand{\lit}[1]{Ref.~\mbox{[\!\!\citenum{#1}]}\xspace}
\newcommand{\lits}[1]{Refs.~\mbox{[\!\!\citenum{#1}]}\xspace}

\definecolor{CBdblue}{RGB}{5,113,176}

\newcommand{\sifig}[1]{Fig.~S{#1}}
\newcommand{\sisec}[1]{Section~S{#1}}

\if\USEACHEMSO0
\begin{document}
\fi

\title{Simulating real-time molecular electron dynamics efficiently using the time-dependent density matrix renormalization group}
\author{Imam S. Wahyutama}
\author{Henrik R.~Larsson}
\affiliation{Department of Chemistry and Biochemistry, University of California, Merced, CA 95343, USA}
\if\USEACHEMSO1
\email{p_tddmrg24 [a t] larsson-research . de}
\else
\email{p_tddmrg24 [a t] larsson-research . de}
\fi

\if\USEACHEMSO1
\begin{document}
\fi

\begin{abstract}
Compared to ground state electronic structure optimizations,
accurate simulations of molecular real-time electron dynamics are usually much more difficult to perform.
To simulate electron dynamics, the time-dependent density matrix renormalization group (TDDMRG) has been shown to offer an attractive compromise between accuracy and cost.
However, many simulation parameters significantly affect the quality and efficiency of a TDDMRG simulation.
So far, 
it is unclear whether common wisdom from ground state DMRG carries over to the TDDMRG,
and a guideline on how to choose these parameters is missing. 
Here, in order to establish such a guideline, we investigate the convergence behavior of the main TDDMRG simulation parameters, such as time integrator, the choice of orbitals, and the choice of matrix product state representation for complex-valued non-singlet states.
In addition, we propose a method to select orbitals that are tailored to optimize the dynamics.
Lastly, we showcase the TDDMRG by applying it to charge migration ionization dynamics in furfural, where we reveal a rapid conversion from an ionized state with a $\sigma$ character to one with a $\pi$ character within less than a femtosecond.
\end{abstract}

\maketitle

\section{Introduction}
\label{sec:intro}

The study of electron dynamics in molecules has become an active research field for the past three decades. \cite{cm-cederbaum-1999, cm-kuleff-2014, cm-phenylalanine-2014, cm-iodo-2015, chemdyn-perspective-2018, quantum-chem-atto-2020, td-elstru-rev-2020}
This trend has been driven by the development of intense ultrashort pulsed laser systems with wavelengths ranging from as short as those in the X-ray regime up to the infrared.\cite{opcpa-1997, xfel-2010, opcpa1-2012}
Among others, electron dynamics enables the control of chemical reactions or even real-time control of the properties of matter,\cite{reaction-control-tannor1-1985, reaction-control-tannor2-1986, select-dissoc-2001, pfeiffer-2015}
and a deeper understanding of biomolecules. \cite{cm-phenylalanine-2014, cm-fmartin-2016, chemdyn-perspective-2018, cm-amino-2019, cm-adc-chordiya-2023}
Alongside the fast progress of experimental methods for ultrafast electron dynamics, accurate theoretical methods that help to interpret experimental observations would be much desirable.
In fact, to tackle ultrafast real-time molecular electron dynamics,
progress has been made to extend and apply concepts from traditional quantum chemistry.
This results in a number of time-dependent extension\rladd{s} of several quantum mechanical methods such as
density functional theory (DFT), \cite{tddft0-1984, tddft-lda-1996, tddft1-2005, tddft-fast-el-2011, tddft3-2011, tddft-isborn-2016}
algebraic diagrammatic construction (ADC), \cite{kuleff-multie-prop-2005, tdadc-ion-hhg-2014, cm-control-laser-2015}
(multireference) configuration interaction (CI), \cite{tdci1-2006, tdcis1-2010, tdrasci-hochstuhl-2012, ionization-tdci2-2014, tdgasci-2014, gasci-2016, larsson-sfi-2016, evangelista-tdaci-2019, ionization-tdci1-2022, tdci-floquet-2022, tdcis2-2022}
multiconfiguration self-consistent field (MCSCF)\cite{tdcasscf1-2013, tdrasscf-2013, tdormas-2015, tdcasscf-application-2016, tdmcscf-resonant-2019}
(also known as multi-configuration time-dependent Hartree-Fock, MCTDHF\cite{zanghellini-mctdhf-2002, mctdhf-kato-2004, mctdhf-caillat-2005, mctdhf-saalfrank-2005, mctdh-book-2009, hochstuhl-mctdhf-2011, hochstuhl-mctdhf-eur-2014, mctdhf-mccurdy-2015, lode-mctdhf-2020})
coupled cluster, \cite{td-occ1-2018, td-occ2-2020, td-occ3-2021, tdcc-2023}
density matrix embedding theory,\cite{kretchmer-dmet-2018}
and the density matrix renormalization group (DMRG),
\cite{tst1-2005, tdvp-lattice-2011, tdvp-tensor_train-2013, tdvp-time_integration-2015, tdvp-unify-2016, ddmrg++-2017, PAECKEL_tddmrg_review-2019, tddmrg-pfannkuche-2019, tddmrg-baiardi-2021, hrl-mctdh-rev-2024}
to name a few.

The prevailing challenge common to many of these methods is the well-known exponential growth of the computational load as the system size increases. Among these methods, the Kohn-Sham real-time time-dependent DFT (RT-TDDFT) \cite{tddft1-2005, tddft3-2011, tddft-isborn-2016}
can be considered to be the most attractive in terms of computational speed, owing to its underlying assumption in the form of a single electronic configuration.
Numerous works employed RT-TDDFT to study the ultrafast electronic motion in molecules that would have been too prohibitive for wave function-based methods.
RT-TDDFT studies have
been able to uncover interesting observations. \cite{tddft-strong-field-2001, vib-branch-2012, tddft-uracil-2014, cm-tddft3-2017, quantum-chem-atto-2020, cm-tddft2-2021, cm-soliton-2022, cm-tddft1-2022, cm-attochem-2023, tddft-coulomb-decay-2024}
The main drawbacks of RT-TDDFT, however,
are its approximate nature and the difficulty of systematically improving it.
This makes it difficult to predict the accuracy of RT-TDDFT results, in particular when the dynamics drives the system far from the ground state. \cite{tddft-error-2003, tddft-critique-2007, cm-single-det-2009, rabi-tddft-2014, td-elstru-rev-2020, tddft-size-error-2023}

In contrast to RT-TDDFT,
the accuracy of
wave function-based %
methods
is largely predictable in the physical regime they are based on (single-reference vs.~multireference scenarios) and often they are systematically improvable.
The price to pay, however, is the high computational cost, making their applicability, %
limited to molecules consisting of only a few atoms or having only a few tens of orbitals. \rladd{Some examples of systems that have been studied using time-dependent wave function-based methods include uracil ($58$ electrons and $272$ orbitals) using single-reference time-dependent ADC(3), \cite{cm-adc-chordiya-2023} explicit charge exchange and ionization dynamics of \ce{BC^2+} with $9$ electrons in $90$ determinants and resonant high-harmonic generation in \ce{Mo} using a CAS($14$e, $10$o) and MCTDHF/time-dependent complete active space self-consistent field (CASSCF), \cite{mctdhf-mccurdy-2015,mo_rhhg-2020} and decacene using a CAS($12$e,$12$o) and 
time-dependent complete active space configuration interaction (TD-CASCI).\cite{levine-tdcasci-2018}}

Time-dependent extensions of the DMRG \cite{dmrg-white-formulation-1992, dmrg-white-algorithms-1993, dmrg_large_scale-2004, dmrg_chan_inbook-2008, dmrg_for_qc-2014, dmrg_in_practice-2015, dmrg_large_casscf-2017, baiardi_recent_dev-2020, dmrg-large-site-2022} (TDDMRG\footnote{Note that different algorithms exist that exploit DMRG-like approaches for time evolution. Here, we use ``TDDMRG'' to generically refer to all of them.})\rladd{\cite{tst1-2005, tdvp-lattice-2011, tdvp-tensor_train-2013, tdvp-time_integration-2015, tdvp-unify-2016, ddmrg++-2017, PAECKEL_tddmrg_review-2019, tddmrg-pfannkuche-2019, tddmrg-baiardi-2021, hrl-mctdh-rev-2024}}
offer an attractive compromise---they are based on first principles and allow for increasing the number of orbitals up to a region that would be off-limit for many multireference wave function-based methods. %
A few studies that demonstrate the capabilities of the DMRG to simulate molecular electron dynamics have been published; among others, these include ionization dynamics, linear response properties, \cite{dmrg-lrt-2014} and charge migration.\cite{tddmrg-pfannkuche-2019, tddmrg-baiardi-2021} %
These works included up to $36$ orbitals.

Due to its efficiency, in particular for multireference scenarios common in nonequilibrium electron dynamics, we believe that the TDDMRG will gain more attention and interest as a method for studying ultrafast electron dynamics in the near future.
However, while ground state DMRG for quantum chemistry by now is established and can (almost) be used in a black-box way, \cite{dmrg-acenes-2007, dmrg_in_practice-2015, baiardi_recent_dev-2020, larsson-cr2-2022}
time-dependent versions of the DMRG are more complicated.
Notably, even the way to solve the time-dependent Schrödinger equation using DMRG-like procedures is done differently in different works.\cite{tst1-2005, tdvp-unify-2016, PAECKEL_tddmrg_review-2019, tddmrg-pfannkuche-2019, tddmrg-baiardi-2021}
Next to the many variants of the TDDMRG, it is unclear how to best choose simulation parameters and whether concepts that work well for the ground state DMRG also apply to the TDDMRG.
As standard DMRG approaches are based on the complete active space (CAS) concept, selecting the appropriate set of orbitals to simulate dynamics is another major issue that so far has not been fully addressed.

Here, we make an attempt toward a better understanding of many of the required ingredients of time-dependent DMRG simulations for molecular electron dynamics.
Our goal is twofold: (1) benchmarking and understanding TDDMRG simulation parameters  and (2) extending procedures from ground state DMRG and CAS-based methods for selecting the most suitable orbitals for the active space to dynamics.

Since the interplay of all of the algorithms and choices that make up an accurate and efficient DMRG simulation is very complex, this work clearly needs to be limited. In particular, we limit ourselves to only two related TDDMRG variants, namely the time step targeting (TST) method and a projector-splitting approximation of the time-dependent variational principle (TDVP).

Furthermore, we have limited the scope of this paper by only studying the dynamics of one particular physical effects,
namely charge migration, which plays a key role in the photofragmentation of peptides \cite{charge-transfer-weinkauf-1996, levine-select-react-1997, charge-direct-react-1998, control-select-reaction-1999, peptide-fragment-pathways-2005, cm-peptide-2007, uv-photogragment-2009}
and has the potential to enable the control of chemical reactions. \cite{cm-select-reaction-covion-1999, reaction-control-kling-2013, cm-control-laser-2015, control-smooth-pulse-2017}
Charge migration is initiated through ionization, where
a positively charged region, called \textit{hole}, migrates across the molecule.
Notably, this effect happens on an ultrafast scale of attoseconds to few femtoseconds, \cite{cederbaum-atto-response-2005, cm-phenylalanine-2014, cm-iodo-2015}
and is mainly driven by correlation. \cite{cm-cederbaum-1999, cm-breidbach-2003, cm-breidbach-num-2007, cm-kuleff-2014}
Thus, its simulation may serve as a measure of how accurate a method is in describing electronic correlation.
Several existing time-dependent methods have been used to simulate ultrafast charge migration; these include ADC, \cite{cm-breidbach-2003, kuleff-multie-prop-2005, cm-breidbach-num-2007, cm-kuleff-2014, cm-control-laser-2015, cm-adc-chordiya-2023}
RT-TDDFT, \cite{cm-tddft3-2017, cm-tddft2-2021, cm-tddft1-2022}
time-dependent CI methods, \cite{evangelista-tdaci-2019, rhodyn-2022, 5-heterocyclic-cm-2023, cm-tdci-2023}
and TDDMRG. \cite{tddmrg-pfannkuche-2019, tddmrg-baiardi-2021}

The outline of the remainder of this paper is as follows: \autoref{sec:theory} gives a brief overview of DMRG ingredients.
\autoref{sec:benchmark} benchmarks various aspects of the TDDMRG, and  \autoref{sec:spatial_orbital} introduces and benchmarks two approaches on how to choose orbitals that are adapted to the actual dynamics of interest.
In \autoref{sec:appl} we apply our findings through an application of charge migration in furfural.
We conclude in \autoref{sec:conclusions}. %

\section{Theory}
\label{sec:theory}
Here, we give \rladd{a concise} overview of some necessary ingredients of TDDMRG simulations. 
\rladd{For more details, we refer to the references cited in the following sections.}
In \autoref{sec:mrci} we briefly discuss time-dependent multiconfiguration-based methods, in \autoref{sec:mps} we  review matrix product states, and in \autoref{sec:dmrg}  the DMRG for computing ground states.
We discuss two ways of propagating a matrix product state (MPS) in \autoref{sec:tdvp} and \autoref{sec:tst}, respectively.
This is followed by an overview of how to deal with
symmetries in an MPS in \autoref{sec:symmetry_qn}, and two ways to describe a complex-valued state by an MPS  in \autoref{sec:complex}.
Lastly, we discuss two ways to make DMRG simulations more efficient, namely the orbital ordering in \autoref{sec:ordering} and general orbital rotations in \autoref{sec:orbital_shape}. %

\subsection{Time-dependent multiconfiguration methods}
\label{sec:mrci}

In general,
time-dependent multiconfiguration methods are based on a linear expansion of the time-dependent state $\ket{\Psi(t)}$ in a basis of configurations $\ket{\Phi_i}$     %
as \cite{multireference-excited-2018}
\begin{gather}
    \ket{\Psi(t)}
    =
    \sum_i
    C_i(t) \,
    \ket{\Phi_i},
    \quad C_i \in \mathbb C.
    \label{eq:gen_tdwfn}
\end{gather}
Multiconfigurational methods differ in the choice of the configurations being used.
For example,
in the \rlremove{time-dependent complete active space configuration interaction (TD-CASCI)} \rladd{TD-CASCI} method, all possible configurations within an active orbital space are included and do not vary with time. \cite{tdrasci-hochstuhl-2012, levine-tdcasci-2018, tdci-floquet-2022}
As the exponential  scaling of the number of possible configurations with respect to the number of active orbitals limits CASCI to ${\sim}20$ orbitals without additional approximations, TD-CASCI mostly captures static electronic correlation.
To capture dynamic correlation, in time-dependent multi-reference configuration interaction (TD-MRCI), configurations with electronic excitations out of the active space are added to the expansion.
To increase the flexibility of \autoref{eq:gen_tdwfn} and hence to  decrease the number of required configurations, these configurations can be made time-dependent by solving Dirac-Frenkel's time-dependent variational principle \cite{lubich-from-quantum-2008}
for both the orbitals and the coefficients $C_i(t)$. 
\cite{mctdh-2000, mctdh-book-2009, hochstuhl-mctdhf-eur-2014, lode-mctdhf-2020} 
For TD-CASCI this leads to the MCTDHF method,\cite{zanghellini-mctdhf-2002, mctdhf-kato-2004, mctdhf-caillat-2005, mctdhf-saalfrank-2005, mctdh-book-2009, hochstuhl-mctdhf-2011, hochstuhl-mctdhf-eur-2014, mctdhf-mccurdy-2015, lode-mctdhf-2020}
and for TD-MRCI this leads to the time-dependent restricted active space self-consistent field (TD-RASSCF) method, \cite{tdrasscf-2013, mctdhf-mccurdy-2015, tdormas-2015} among others.
While MCTDHF and related methods require much smaller orbital spaces, compared to TD-CASCI, the implementation is more complicated and the method can be numerically challenging. \cite{mctdh-revisit-2015, mctdh-regularize-2018}
\subsection{Matrix product states\label{sec:mps}}

MPSs approximate the CAS-CI coefficients $C_i$ in \autoref{eq:gen_tdwfn} as products of matrices $\mathbf M^{\sigma_i}$:
\begin{gather}
    \ket{\Psi} = \sum_{\{\sigma\}} \left(\prod_{i=1}^K \mathbf M^{\sigma_i} \right)
    \ket{\sigma_1 \ldots \sigma_K},
    \label{eq:mps_form}
\end{gather}
where $K$ is the number of spatial orbitals, each of which resides on an MPS \textit{site} (hence, there are $K$ sites).  The physical basis states $\{\ket{\sigma_i}\}$ %
consist of all possible spin occupancies of the $i$-th spatial orbital $\ket{\phi_i}$, that is $\ket{\sigma_i} \in \{\ket{\text{vac}}, \ket{\phi_i^\alpha}, \ket{\phi_i^\beta}, \ket{\phi_i^{\alpha\beta}} \}$. For simplicity, here we have assumed no spin adaptation (cf. \autoref{sec:symmetry_qn}).
The matrices $\mathbf M^{\sigma_i}$ are of size $D_{i-1}\times D_{i}$, save for the first and last ones, which are row and column vectors, respectively. %
The \emph{bond dimension} is then defined as $D=\max_{i} D_i$. The larger $D$ is, the more accurate the MPS approximation will be. Since $\sigma_i$ has four possible values, \mpsm{i} can be viewed as a slice of a three-dimensional tensor \mpst{i} of size $4 \times D_{i-1} \times D_i$.
Depending on the context, in the following we will use the same symbol $\mpst{i}$ also to denote a vectorized tensor.

The matrices appearing in \autoref{eq:mps_form} are not unique, since one can insert an identity matrix $\mathbf{I} = \mathbf X^{-1} \mathbf X$ in between any pair of matrices without changing $\ket{\Psi}$. This gauge freedom can be exploited to impose an orthonormality condition on $\mpsm{i}$. If it satisfies $\sum_{\sigma_i} (\mpsm{i})^\dagger \mpsm{i} = \matr I$, $\mpsm{i} \rightarrow \mpsml{i}$ and is called left-orthogonalized. If it satisfies $\sum_{\sigma_i} \mpsm{i} (\mpsm{i})^\dagger = \matr I$, $\mpsm{i} \rightarrow \mpsmr{i}$ and is called right-orthogonalized.
Using $I-1$ left-orthogonalized sites, matrices $\mathbf M^{\sigma_I}$ at site $I$, and $K-I-1$ right-orthogonalized sites leads
to %
\begin{gather}
    \ket{\Psi} = \sum_{\{\sigma\}} \left(\prod_{i=1}^{I-1} \mathbf L^{\sigma_i} \right)
    \mathbf M^{\sigma_I}
    \left(\prod_{i=I+1}^{K} \mathbf R^{\sigma_i} \right)
    \ket{\sigma_1 \ldots \sigma_K}.\label{eq:orthform}
\end{gather}
Then the norm of the vectorized tensor $\mathbf M^{[I]}$ is  identical to that of the whole MPS.
For the form in \autoref{eq:orthform}, the site $i=I$ is called orthogonality center.
Introducing so-called  renormalized states
 $\ket{\lstatei{I}}$ and $\ket{\rstatei{I}}$
of a subsystem formed by the first $I$ sites and the last $K-I-1$ sites, respectively, simplifies \autoref{eq:orthform} to
\begin{gather}
    \ket{\Psi} = \sum_{\alpha_{I-1} \sigma_I \alpha_{I}}
    \mpsm{I}_{\alpha_{I-1} \alpha_{I}}
    \ket{\lstatei{I-1} \sigma_I \rstatei{I}}.
    \label{eq:renorm_expand}
\end{gather}
In any given MPS, the orthogonality center can be shifted to the previous or next site by performing a QR decomposition on either $(\mpst{I})^T$ or $\mpst{I}$, e.g.,
\begin{equation}
\mpst{I} = \mathbf L^{[I]}\mathbf C^{[I]},
\label{eq:qr_decomp}
\end{equation}
where the matrix $\mathbf C^{[I]}$ is then absorbed in $\mathbf R^{[I+1]}$ to obtain the new orthogonality center $\matr M^{[I+1]}=\mathbf C^{[I]}\mathbf R^{[I+1]}$.
\subsection{Density matrix renormalization group\label{sec:dmrg}}

To solve the time-independent Schrödinger equation,
in the DMRG algorithm, the MPS is variationally optimized by
freezing all but the orthogonality center. This then leads to an eigenvalue problem for an effective Hamiltonian matrix $\matr H^{[I]}$ that is represented in the renormalized and physical basis shown in \autoref{eq:renorm_expand}:
\begin{gather}
    \mpsg{H}{I} \mpst{I} = E \mpst{I}
    \label{eq:eigenvalue_problem}
\end{gather}
The ground state vector of  $\matr H^{[I]}$ thus leads to an improved tensor $\matr M^{[I]}$.
Based on this, the orthogonalization center is successively changed from the first to the last site,
and at each site $I$, the effective Hamiltonian is diagonalized and  $\matr M^{[I]}$ is updated.
This procedure is called \emph{sweep}.
Repeating this for multiple sweeps leads to a fully optimized ground state MPS.
Note that typically, the orbitals are not optimized, so the standard DMRG only optimizes the MPS that approximates the CAS-CI coefficients. DMRG optimization together with orbital optimization, as done in CASSCF, is possible as extension and called here DMRGSCF. \cite{dmrgscf-ghosh-2008, dmrgscf-nooijen-2008}

The sweep procedure outlined in the previous paragraph amounts to the so-called one-site DMRG, which is known to be prone to getting stuck at local minima. To overcome this limitation,
in two-site DMRG, a ``two-site tensor'' $\mpst{I,I+1} = \mpst{I} \mpstr{I+1}$ is constructed and the effective Hamiltonian of the corresponding renormalized basis is solved as in one-site DMRG.
After $\mpst{I,I+1}$ is replaced with the ground state of the effective Hamiltonian,
to again obtain two sites $\mpst{I}$ and $\mpstr{I+1}$, a singular value decomposition or similar type of decomposition of  $\mpst{I,I+1}$ is performed.
The remaining steps are analogous to the one-site version.
\subsection{Time-dependent variational principle \label{sec:tdvp}}
A DMRG-type of approximation of the time-dependent variational principle (TDVP) leads to an algorithm that is very similar to the DMRG.\cite{tdvp-time_integration-2015, tdvp-unify-2016}
Note that here we will not consider orbital optimization, i.e., we approximate TD-CASCI using the DMRG and use time-independent orbitals.
In the TDVP variant of the TDDMRG, the diagonalization of the effective Hamiltonian $\matr H^{[I]}$ in DMRG, is replaced by a propagation by a time step $\Delta t/2$:
\begin{align}
    \mpst{I}\left( t + \Delta t/2 \right) =&\,
    \exp\left[ -i \mpsg{H}{I} \Delta t/(2\hbar) \right] \mpst{I}(t),
    \label{eq:tdvp_step1}
\end{align}
To move the orthogonalization center, the QR decomposition  from \autoref{eq:qr_decomp} is performed, giving $\mpst{I}\left( t + \Delta t/2 \right) = \mathbf L^{[I]}\left( t + \Delta t/2 \right) \mathbf C^{[I]}\left( t + \Delta t/2 \right)$.
Before $\matr{C}^{[I]}(t+\Delta t/2)$ can be absorbed into  $\matr{R}^{[I+1]}(t)$ to obtain the new orthogonalization center $\mpst{I}(t)$, the two propagation times of the tensors need to be aligned. This is achieved by backward-propagating $\matr{C}^{[I]}(t+\Delta t/2)$ to time $t$ using \autoref{eq:tdvp_step1} but with a negative time step.
This backward propagation is the only part of the TDVP version of the TDDMRG that does not have a counterpart in the DMRG.
\rladd{Note that the renormalized basis used for backward-propagating $\matr{C}^{[I]}(t+\Delta t/2)$ is different from that used for forward-propagating $\mpst{I}(t)$. %
This is one of the error sources in TDVP, which can be minimized by decreasing the time step.}
Performing \rlremove{this propagation} \rladd{this sequence of forward and backward propagations} for sites $1$ to $K$ and then backward for sites $K$ to $1$ propagates the total MPS from time $t$ to $t+\Delta t$ with an error of $\mathcal O(\Delta t^3)$.
As in the DMRG, a two-site version can be introduced as well.
Since this algorithm is based on the TDVP, it is often dubbed TDVP-DMRG.
It is also known as the projector splitting integrator in the multi-layer MCTDH community, where this algorithm is used to propagate tree tensor networks states, an extension of MPSs. \cite{tdvp-tensor_train-2013, lubich-mctdh-integrate-2015, tdvp-time_integration-2015, psi-novel-2017, mctdh-tevo-appl-2021, hrl-mctdh-rev-2024}
For more details, we refer to Refs.\citenum{tdvp-time_integration-2015, tdvp-unify-2016, PAECKEL_tddmrg_review-2019, hrl-mctdh-rev-2024}.
\rladd{As with many other propagation methods, eigenstates can be obtained by propagating in imaginary time. Indeed, in the limit of $\Delta t \to -i\infty$, the DMRG algorithm for eigenstates is recovered from the TDVP algorithm.\cite{ tdvp-unify-2016}}

Since the time step is a convergence parameter, how can we estimate a reasonable time step?
An estimate based on the time-energy uncertainty relation frequently provides a reasonable guess:\cite{tdse-compare-1991} %
\begin{equation}
  \Delta t_E \sim  \frac{\hbar}{\Delta E} = \frac{\hbar}{E_\text{max} - E_\text{min}},
  \label{eq:time_step_estimate}
\end{equation}
 where $E_\text{min}$ ($E_\text{max}$) are the ground state (highest) energy of the Hamiltonian.
Using the DMRG, $E_\text{max}$ can be computed by solving for the ground state of $-\hat H$.

\subsection{Time-step targeting \label{sec:tst}}
In
time step targeting (TST),\cite{tst1-2005}
the central idea for solving the time-dependent Schrödinger equation is to optimize a renormalized basis at each time step $\Delta t$ that describes the evolution from $\ket{\Psi(t)}$ to $\ket{\Psi(t+\Delta t)}$.
In practice, this is done by a sweep algorithm and a fourth-order Runge-Kutta propagator, albeit other propagators are possible as well.\cite{te-mps-ripol-2006, PAECKEL_tddmrg_review-2019}
In the sweep algorithm, at each site, four Runge-Kutta vectors are generated through four matrix-vector products with the effective Hamiltonian. From these four vectors a new renormalized basis is created in a state-average fashion. %
This can be repeated for a pre-defined number of sweeps. 
\rladd{Since the state-averaged intermediate state requires a larger bond dimension than $\ket{\Psi(t)}$ alone, sweeping optimizes the state-averaged renormalized basis.}
In a final sweep, the wave function at time $t+\Delta t$ is updated using the renormalized basis created through the previous sweeps.
Note that TST is non-unitary even in the one-site variant.
While TDVP can be regarded as more rigorous than TST,\cite{PAECKEL_tddmrg_review-2019} TST is still used in daily research. \cite{conserve-cc-ruojing-2021, block2-2023, fermion-mps-2024}

Next to TST and TDVP, many other MPS propagation methods exist.\cite{out-equil-mps-2012, PAECKEL_tddmrg_review-2019}
While most of them such as  time-evolving block decimation are more targeted to systems with short-range interactions and not fully suitable for molecules, some other methods can also be used for molecular electron dynamics. In particular,  the global Lanczos method has been used previously for charge migration.\cite{tddmrg-pfannkuche-2019}
Therein, the short iterative Lanczos (SIL) propagator \cite{sil-light-1986, tannor-book-2007} is used for the calculation of the action of a matrix exponential on a vector, which here is given by the MPS.
The necessary applications of the Hamiltonian onto the global MPS as well as MPS linear combinations increase the bond dimension at every step of the SIL algorithm. Thus, additional MPS compressions are required and lead to non-unitary dynamics.

\subsection{Exploiting symmetries\label{sec:symmetry_qn}}

For the field-free dynamics considered here, electron number, spin, and molecular point group symmetries are conserved.
To impose Abelian symmetry such as electron number, $z$ component of the total spin, and point group symmetry, one enforces symmetry conditions on the coefficient tensor $\matr M^{[I]}$ in \autoref{eq:renorm_expand}
such that the quantum numbers of the combined left renormalized basis $\{\ket{l_{\alpha_i}}\}$ and physical basis
$\{\ket{\sigma_i}\}$ match those of the right renormalized basis $\{\ket{r_{\alpha_i}}\}$. \cite{tn-anthology-2019}
Imposing this condition across the entire MPS leads to a block-sparse structure of the MPS tensors $\mpst{i}$.

The imposition of the non-Abelian total spin symmetry is more involved than Abelian symmetries, as more than one pair of total spins of the renormalized bases can yield the same global spin. %
The global spin symmetry can be imposed by decomposing each site tensor into two tensors, the first one containing the Clebsch-Gordan coefficients, and the second one containing the rotation coefficients that form the renormalized states for that site. \cite{dmrg-spin-symm-2012, dmrg_for_qc-2014}
Alternatively, one can also use projectors to implement Abelian and non-Abelian symmetries, resulting in a simpler albeit less efficient implementation. \cite{spin-project-mps-2017, mmps-larsson-2020}

Spin-adapted MPSs for non-singlet states are more involved than singlet states, as many more spin couplings are possible.
To avoid this, singlet-embedding can be used. \cite{tatsuaki-singlet-embed-2000, dmrg-spin-symm-2012}
Therein, one adds non-interacting auxiliary orbitals at the end of the MPS. These orbitals couple to the physical orbitals such that the total spin is zero. Since they
do not interact with the Hamiltonian, the smaller target MPS is well-represented. %
\subsection{Representation of complex-valued matrix product states \label{sec:complex}}

For solving the time-dependent Schr\"odinger equation,
complex-valued MPSs are needed, and to represent them, there are three main ways.
In the first way, dubbed here full complex MPS,
all MPS tensors $\matr M^{[i]}$ are complex-valued.
This is the most straightforward way but increases memory requirements by up to a factor of 4.

In the second way,\cite{ddmrg++-2017} $\Re( \ket{\Psi(t)})$ and $\Im (\ket{\Psi(t)})$
are interpreted as two different states and a state-averaged (SA) representation is used to describe both real and imaginary parts by one single MPS, where only the tensor belonging to the orthogonalization center is complex-valued.
Consequently, the movement of the orthogonality center of an MPS in  complex SA format has to be adapted, as a renormalized basis needs to be constructed that describes both the real and imaginary parts of the previous orthogonalization center, before moving to the new center. In practice this can be done by either diagonalizing a state-averaged density matrix constructed from the real-valued density matrices  of $\Re(\matr M^{[I]})$ and $\Im(\matr M^{[I]})$, \cite{ddmrg++-2017}
or through singular value decomposition. \cite{larsson-ttns-vib-2019}
This modified orthogonalization is only exact when the bond dimension is allowed to increase.
In practice, however, the bond dimension is truncated, which renders the orthogonalization of an SA MPS inaccurate.
Hence, while the SA complex format leads to greatly reduced computational and memory requirements, it also introduces an additional truncation error and leads to non-unitary dynamics even for the one-site TDVP.
Nevertheless, this format is used in practice,\cite{conserve-cc-ruojing-2021, block2-2023, fermion-mps-2024}
and it is particularly useful for codes that only support real-valued algebra.

The third way to deal with complex-valued MPSs is to describe the real and the imaginary part of the state separately by two distinct MPSs. This way, however, imposes a severe restriction on the used propagation method and thus is rarely used in practice.

\subsection{Orbital ordering}
\label{sec:ordering}

Due to the linear nature of the MPS, a small perturbation at an orbital/the corresponding site at one end of the MPS needs to ``propagate'' through the entire MPS before the orbital/site at the other end of the MPS can ``respond.'' Thus, the MPS is variant under orbital permutations. To achieve optimal performance,
the orbitals should be arranged such that those that correlate largely with each other are placed close together.
Several orbital ordering methods exist, such as
the  symmetric reverse Cuthill–McKee ordering, \cite{dmrg-study-polynom-2002}
approximate best prefactor ordering using simulated annealing, \cite{sv-inverse-symm-2021}
Fiedler ordering,\cite{fiedler1-2011} and using a genetic algorithm to find a close candidate to the global minimum.\cite{ga-reiher-2004, dmrg_in_practice-2015}
They are based on minimizing auxiliary quantities such as the bandwidth of the exchange integral matrix.
Localized orbitals (see \autoref{sec:orbital_shape}) of molecules with simple geometries such as the linear ones can also be ordered based on the orbitals' centers.
\subsection{Orbital shape}
\label{sec:orbital_shape}
The DMRG is variant under unitary rotations of the orbitals as long as the bond dimension is not converged.
Hence, similar to the ordering of the orbitals,
optimizing the orbital shape is important to reduce the entanglement between the orbitals. \cite{dmrgscf-ghosh-2008, dmrg_in_practice-2015, fermion-orb-optim-2016}
For ground state DMRG simulations of small molecules, natural orbitals are often used, as they decrease the bandwidth of the Hamiltonian and thus allow for ``energy localization.''\cite{natorb-for-dmrg-2013, dmrg_in_practice-2015, larsson-cr2-2022}
For larger molecules, however, compared to natural orbitals,
split localization  in coordinate space
(localizing occupied and virtual spaces separately)
significantly reduces the overlap between orbitals and thus also the entanglement and the required bond dimension. \cite{orbloc1-2012, dmrg_in_practice-2015, dmrg_large_casscf-2017, baiardi_recent_dev-2020, dmrg-large-site-2022}
Compared to a global orbital localization of all orbitals, split localization ensures that low-energy configurations are still well-described by a small-$D$ MPS.
Based on a good initial guess, one can also variationally optimize the orbitals during the DMRG optimization. \cite{dmrgscf-ghosh-2008, dmrgscf-nooijen-2008, fermion-orb-optim-2016}
\section{Benchmarking TDDMRG \label{sec:benchmark}}

In the following,
we study the convergence behavior of TDDMRG with respect to various parameters to  establish a robust simulation framework.
We start with a discussion of the system setup in \autoref{sec:methods}
and of the used observables to measure convergence in  \autoref{sec:conv_measure}. Then in \autoref{sec:time_integration}, we compare the convergence behavior of the TDVP and TST methods.
We  benchmark complex MPS representations in
\autoref{sec:complex_mps} and singlet embedding in  \autoref{sec:singlet}. We finish the benchmark with a comparison of natural orbitals and localized orbitals in \autoref{sec:localization}.

All our simulations are performed using custom scripts that utilize the DMRG program package and library \textsc{block2}, \cite{block2-2023, block2-gh-2024}
together with \textsc{PySCF} for the remaining quantum chemical computations. \cite{pyscf-wires-2018, pyscf-jcp-2020, pyscf-gh-2024}
We compute molecular integrals using \textsc{Libcint}. \cite{libcint-2015, libcint-gh-2024}

\subsection{Setup \label{sec:methods}}

We choose choloracetylene and furan as test beds.
Choloracetylene displays charge migration similar to the well-studied iodoacetylene,\cite{cm-iodo-2015}
and is linear and small enough to allow for the inclusion of all available non-core orbitals of a double-$\zeta$ basis in converged TDDMRG simulations without too large values
for the bond dimension.
Compared to chloroacetylene, furan is a (slightly) larger prototypical heterocyclic
molecule with more complicated charge migration. \cite{5-heterocyclic-cm-2023}
For furan, we use the 6-31G basis set, \cite{6-31g_1-1971, 6-31g_2-1972} resulting in a frozen-core cation full configuration interaction (FCI) space of $48$ and $25$ electrons, while for chloroacetylene we use the def2-SV(P) basis set, \cite{def2-sv_p_-2005} resulting in a frozen-core cation FCI space of $41$ orbitals and $15$ electrons.
Since our simulations do not target core ionization, we use the frozen core approximation, thus freezing seven (five) orbitals in chloroacetylene (furan).
The used geometries are given in \sisec{1} of the Supporting Information (SI).
Note that here, we limit the convergence studies to
a particular Gaussian basis size, as our objective is the convergence of TDDMRG and not that of the basis set.

To prepare the initial state for the charge migration dynamics, we emulate the state produced by a sudden ionization, %
where we apply a linear combination of annihilation operators $\hat a_i$ on the ground state of the neutral molecule $|\Psi_\text{GS}\rangle$ as %
\begin{align}
    |\Psi(t=0)\rangle &= \sum_i c_i \hat a_i |\Psi_\text{GS}\rangle, %
\end{align}
where $|\Psi(t=0)\rangle$ is the initial cationic state for the TDDMRG simulation.
The linear combination of annihilation operators
allows us to describe ionization process out of a specifically constructed orbital $\ket{\psi}$.
The coefficients $c_i$ are then calculated as the overlap with the molecular orbitals $\{\ket{\phi_i}\}_i$ that are used for representing the MPS:
\begin{equation}
    c_i = \langle \phi_i | \psi \rangle,
\end{equation}
followed by normalization, if $|\psi \rangle$ is not fully spanned by $\{ |\phi_i \rangle \}$.
This way, not only can we control the shape of the ionized orbitals but also use orbitals that are optimal for the MPS representation (see \autoref{sec:localization}).
The sudden ionization approximation can be improved by
using specialized methods to calculate ionization rates in many-electron systems, \cite{tr_mewfat-2014, dft-mewfat-2022} which, however, is not necessary for the purpose of our benchmark.

For our choloracetylene simulations, we choose $\ket{\psi}$ to be an in-phase $50$/$50$ superposition of the Hartree-Fock HOMO and HOMO-1. Each of these orbitals is a %
$\pi$ orbital belonging to a $b_1$ irrep of the $C_{2v}$ point group. The resulting orbital $\ket{\psi}$ is a $p$-like atomic orbital localized at the Cl atom.
For our furan simulations, $|\psi \rangle$ is an intrinsic bond orbital (IBO)\cite{ibo-2013} with a $p$-like symmetry localized at the O atom.
All of the initial states described above transform according to a particular irrep of the point group of the molecule, and, since the dynamics is field-free, the time evolution conserves the irrep of the initial state. Hence, we make use of the point group symmetry. For chloroacetylene we use the $C_{2v}$ point group, and for furan we use $C_s$. %

Our simulations assume frozen nuclei.
Since, in reality, after a few femtoseconds the nuclear motion will start to influence the dynamics,  we run all simulations presented here for $\unit[80]{a.u.}\approx\unit[2]{fs}$ only.
As will be shown in \autoref{sec:conv_measure}, %
this time window already captures important dynamics of the evolving electronic density.

The simulations for chloroacetylene use a CAS with 30 orbitals that are based on the hole-DM adaption procedure described in \autoref{sec:spatial_orbital} and benchmarked in \autoref{sec:dyn-adapt}.
For results other than those shown in \autoref{sec:localization},
all simulations are based on the split-localized orbitals described in \autoref{sec:localization}. %
The simulations for furan in \autoref{sec:localization} employ a CAS with 30 orbitals obtained from the  first 30 non-core \rladd{second-order Møller-Plesset theory (MP2)} natural orbitals.
Throughout, the initial state has the same bond dimension as that used for the dynamics, except for $D=1200$ in chloroacetylene, where the initial state has a bond dimension of $D=1000$, as its energy is already converged to $\myunit[0.07]{\text{m} E_H}$.
For all simulations shown below, we use the TDVP method
with a time step of $\Delta t=\unit[0.968]{as}$, and singlet embedding,  unless stated otherwise.
For the TDVP time propagation in \autoref{eq:tdvp_step1}, we use the SIL method, \cite{sil-light-1986, tannor-book-2007} as implemented in \textsc{block2}. The implementation is  based on \textsc{expokit}. \cite{expokit-1998}
This propagator is based on an adaptive time step with a relative convergence tolerance set to  $5\times 10^{-6}$. A convergence study of this tolerance parameter is shown in \sifig{3} in the SI.
All simulations use the two-site TDDMRG variant.
Since two-site simulations are not norm-conserving, we 
\rladd{calculate the expectation value of any operator $\hat O$ as $\matrixe{\Psi}{\hat O}{\Psi} / \braket{\Psi}{\Psi}$.}
\rlremove{take the loss of norm into account when computing expectation values.}
\subsection{Convergence measures\label{sec:conv_measure}}

As numerical methods converge differently for different observables, as possible convergence criteria,
here we consider three quantities.
They are the hole density, the autocorrelation function, and the
L\"owdin partial charge. We will choose the most sensitive one for most of our comparisons.
The hole density $h(\mathbf r,t)$
is defined as \cite{cm-cederbaum-1999}
\begin{equation}
    h(\mathbf r, t) = \rho_0(\mathbf r) - \rho(\mathbf r, t),
    \label{eq:hole_dens}
\end{equation}
where $\rho_0(\mathbf r)$ and $\rho(\mathbf r, t)$ are the one-particle reduced densities of the ground state neutral molecule and that of the evolving cation, respectively.
The  autocorrelation function is defined as
\begin{gather}
    A(t) = \braket{\Psi(0)}{\Psi(t)}.
    \label{eq:autocorrelation}
\end{gather}
The L\"owdin partial charge $Q_A$ around nucleus $A$ is defined as \cite{szabo1996modern-1996}
\begin{gather}
    Q_A = Z_A - e \sum_{i \in A} P_{ii},
    \label{eq:define_lowdin}
\end{gather}
where $Z_A$ is the charge of nucleus $A$ and $P_{ii}$ is the diagonal matrix elements of the spin-summed reduced density matrix
represented by Löwdin-orthogonalized atomic orbitals. \cite{lowdin-non-ortho-1950}

\begin{figure*}[!htbp]
    \includegraphics[width=\figfull]{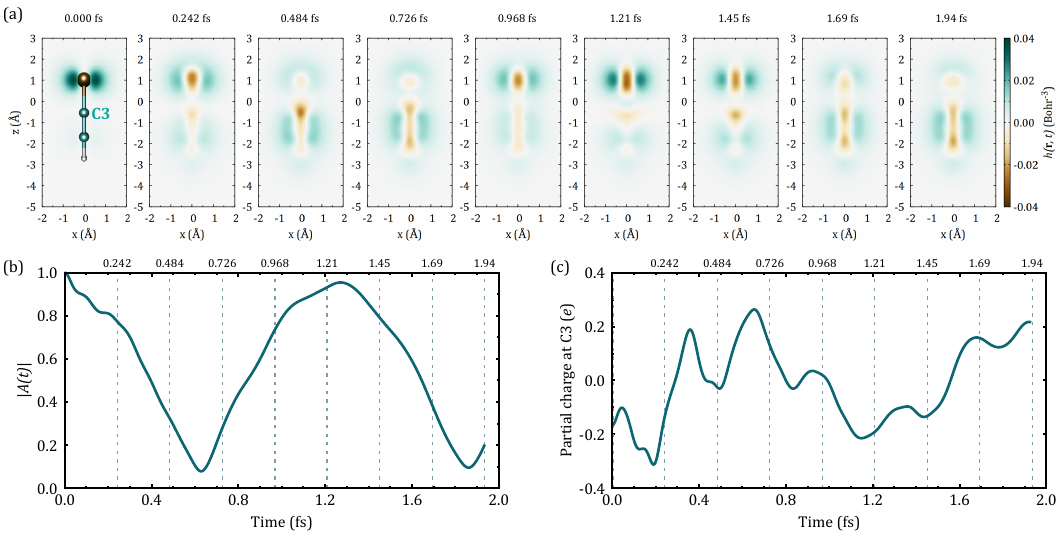}
    \caption{Example of TDDMRG simulations of chloroacetylene. (a) Slices of the hole density $h(\mathbf r, t)$ sampled at nine time points. The slices are in-plane with the molecule and are parallel to the lobes of the initial hole, as can be confirmed from the shape of the %
    region of $h(\mathbf r, t)$ in the leftmost panel.
    (b) Absolute value of the autocorrelation function as function of time.
    (c) $Q_\text{C3}(t)$, the partial charge at C3 (the carbon atom adjacent to Cl; see the molecule in panel (a)) as a function of time.
    The vertical lines in panels (b) and (c) mark the times at which the hole densities in panel (a) are evaluated. For this simulation, the bond dimension is $1000$.
    }
    \label{fig:showcase}
\end{figure*}

Examples of these three time-dependent observables---hole density, autocorrelation function, and L\"owdin partial charge---are shown in \autoref{fig:showcase} for charge migration in chloroacetylene.
The time-dependent hole densities shown as slices in \autoref{fig:showcase}(a) are very useful for capturing the main features of the charge migration dynamics and typically converge quickly.
The autocorrelation function shown in \autoref{fig:showcase}(b)
tracks the contribution of the initial state component in the evolving state, and, for this particular dynamics, displays large revivals and a relatively simple structure.
\rladd{The apparent revival period of  ${\sim}\unit[1.29]{fs}$ is due to the creation of the initial state, which is dominated by a superposition of the ground and the excited states of the cation.
These eigenstates are ${\sim}\unit[3.2]{eV}$ apart in energy, thus resulting  in the observed revival period. Due to electronic correlation and non-Koopman relaxation effects, the actual autocorrelation function is modulated and more structured.
}
Compared to hole densities and the autocorrelation function, however, time-dependent partial  charges shown in \autoref{fig:showcase}(c) contain many frequency components and reveal much more detailed information about the dynamics.
Consequently,
compared to the other two observables,
it is the most sensitive quantity with respect to the parameters of the TDDMRG simulations.
We therefore use partial charges in most of the studies below to judge when convergence is reached by visually inspecting the partial charge plots.

For the chloroacetylene benchmark, we use the partial charge of the carbon atom neighboring the chlorine atom as a measure of convergence, whereas for furan we use the partial charge at one of the carbon atoms on the opposite side of the oxygen atom.

\subsection{Time propagation methods \label{sec:time_integration}}
In the following, we will benchmark the TST and TDVP time propagation algorithms. Using a converged time step, we first inspect the convergence with respect to bond dimension in \autoref{sec:prop_d}. Then, using a converged bond dimension, we inspect the convergence with respect to the time step in \autoref{sec:prop_dt}.

\subsubsection{Convergence of TST and TDVP with respect to bond dimension \label{sec:prop_d}}
\begin{figure}[!tbp]
    \includegraphics[width=\fighalf]{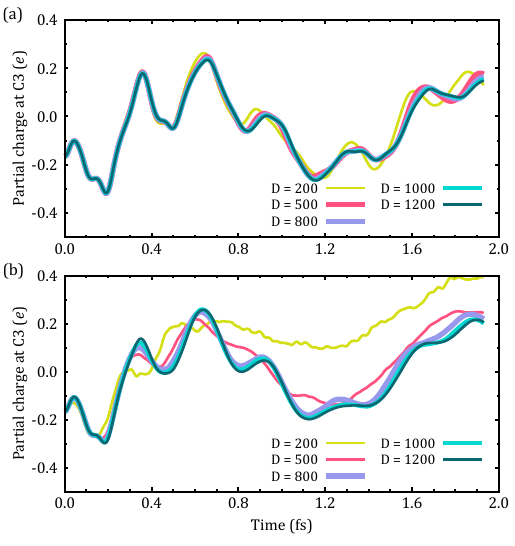}
    \caption{Convergence of partial charge with respect to
    bond dimension for two time propagation methods in choloracetylene.
    (a) $Q_\text{C3}(t)$ obtained using TDVP and (b) using TST.
    The time step is $\unit[0.484]{as}$ and the MPS is of full complex type.}
    \label{fig:prop-pcharge-D}
\end{figure}

\autoref{fig:prop-pcharge-D} shows the bond dimension convergence behavior in chloroacetylene for the TDVP and the TST methods.
Here, both TDVP and TST converge around $D\sim 1000$.
However, TDVP converges smoother and leads to smaller errors for less-converged bond dimensions than TST.
For instance, the $D=500$ TDVP curve in \autoref{fig:prop-pcharge-D}(a) is far closer to the converged curve than the $D=500$ TST curve in \autoref{fig:prop-pcharge-D}(b).
Since, for typical simulations, the required bond dimension increases with time, both for TDVP and TST, the deviations to the converged result increase with time.
As the evolution progresses, simulations with  smaller bond dimensions deviate before those with larger bond dimensions.
For example, using TST, the $D=200$ curve starts to deviate at about $\unit[0.15]{fs}$, while for $D=500$, the deviation occurs at $\unit[0.3]{fs}$.
Compared to TDVP, for a given bond dimension, the point of deviation in the TST dynamics appears earlier.
Remarkably, the $D=200$ TST curve displays a large error and a qualitatively different behavior for times larger than ${\sim}\unit[0.5]{fs}$ whereas the $D=200$ TDVP curve is still qualitatively correct throughout the dynamics, save for some dephasing. A bond dimension convergence scan for another molecule, acetylene, can be found in \sifig{1} in the SI, where the better convergence of TDVP, compared to TST, is retained.
The slower bond dimension convergence of TST, compared to TDVP,
might be due to the state-average-like procedure of the Runge-Kutta vectors, next to other errors not shared with TDVP such as the Runge-Kutta approximation.
The number of sub-sweeps in each time step in our TST simulations is two, except for the first time step in which it is four to allow for a better adjustment of the renormalized states. Results obtained using more sub-sweeps are shown in \sifig{4(a)} of the SI, which, however, do not show improvement in the TST bond dimension convergence.
A different version of TST, where the number of states to average over is decreased by using independent renormalized bases for  $\ket{\Psi(t)}$ and $\ket{\Psi(t+\Delta t)}$\cite{ddmrg++-2017} could decrease the error for small bond dimensions, but that version also requires twice as many resources as the normal TST version.

\subsubsection{Behavior of TDVP and TST with respect to time step \label{sec:prop_dt}}
\begin{figure}[!tbp]
    \includegraphics[width=\fighalf]{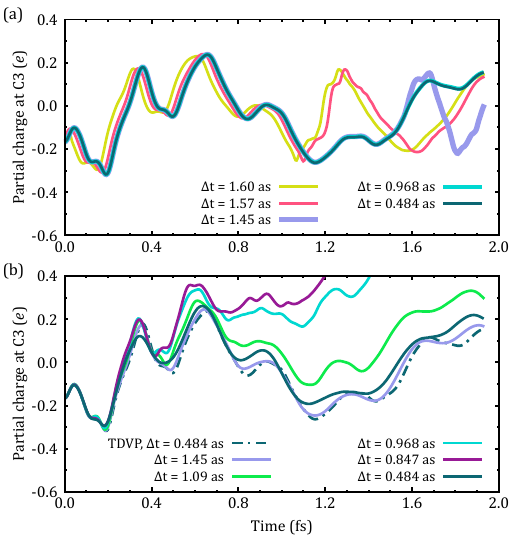}
    \caption{Convergence of partial charge with respect to time step for two time propagation methods in chloroacetylene.
     (a) $Q_\text{C3}(t)$ obtained using TDVP and (b) using TST. To aid in the comparison between TST and TDVP, the curve of $\Delta t = \unit[0.484]{as}$ from panel (a) is also shown in panel (b). The bond dimension is $1000$ and the MPS is of full complex type.
     }
    \label{fig:prop-pcharge-dt}
\end{figure}

Next, we study the convergence of TDVP and TST with respect to the time step. %
\autoref{fig:prop-pcharge-dt} displays the time step convergence in chloroacetylene simulated using TDVP (panel a) and TST (panel b), respectively. The time step convergence of TDVP is smooth and convergence is attained at $\Delta t=\unit[0.968]{as}$.
This is close to the time step estimate from \autoref{eq:time_step_estimate},
which gives $\Delta t_E= \unit[0.656]{as}$.

The time step convergence in TST, however, is not monotonic. At $\Delta t=\unit[1.45]{as}$, the partial charge curve has a good agreement with the converged TDVP curve. However,
as the time step is reduced, the TST curves diverge (see curves for $\Delta t=\unit[1.09,0.968,0.847]{as}$ in \autoref{fig:prop-pcharge-dt}(b)) before they again approach convergence at $\Delta t=\unit[0.484]{as}$.
Note that a smaller time step decreases the Runge-Kutta propagation error in TST but also increases the number of required compressions for the full simulation, resulting in a complex error behavior. \cite{ddmrg++-2017, PAECKEL_tddmrg_review-2019}
Increasing the number of sub-sweeps per time step (two are used here) often reduces the compression error, but doing so does not improve convergence in this case (see \sifig{4(b)} in the SI).
Hence, in this case, TDVP is much more stable against the total time discretization error than TST.
Like the bond-dimension convergence, TDVP curves with larger time step tend to start to  deviate  at earlier propagation times than those with smaller time step, but the convergence is typically smooth. %
As side note, for time steps larger than those shown here, we observed jaggedness in the observables as a function of time when using TDVP.

We note that for simulations  of other molecules (e.g.~acetylene shown in \sifig{2} in the SI), we observed smooth convergence also for TST, albeit TDVP typically converged faster than TST.
Reflecting on the better overall performance of TDVP compared to TST as shown here and in \autoref{sec:prop_d}, we choose to use TDVP for the remainder of this paper.

\subsection{Complex MPS type \label{sec:complex_mps}}
\begin{figure}[!tbp]
    \includegraphics[width=\fighalf]{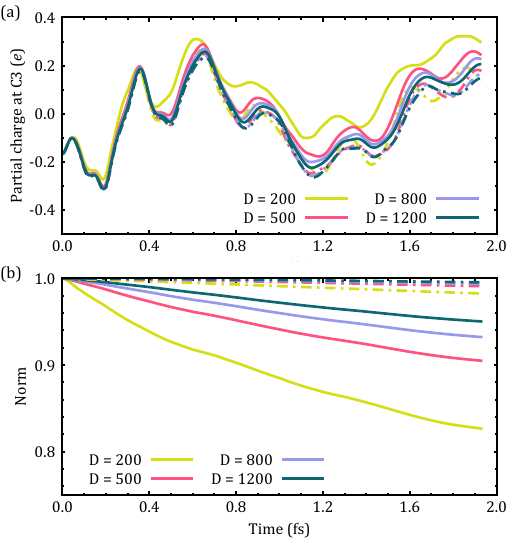}
    \caption{Bond dimension convergence in chloroacetylene for two complex MPS representations. $Q_\text{C3}(t)$ (a)
    and time-dependent norms (b)
    obtained using a complex MPS in state-averaged (solid lines) and in full complex (dot-dashed lines) representations.} %
    \label{fig:complex}
\end{figure}

As explained in \autoref{sec:complex}, a complex-valued MPS can be approximated by a state-averaged version where only the orthogonality center is complex-valued or as a proper complex MPS with all tensors being complex-valued.
For these two types of complex representation,
\autoref{fig:complex}(a) shows the time-dependent partial charge at C3 of chloroacetylene for several bond dimensions.
The full-complex MPS (solid curves) converges with bond dimension noticeably faster than the SA-complex MPS (dot-dashed curves). For example, the $D=200$ full-complex MPS curve is much closer to the reference  ($D=1200$) than the SA complex MPS, which displays significant deviations.

In agreement with the convergence of the partial charge,
observing the norm as a function of time in \autoref{fig:complex}(b), it becomes evident that the SA complex type (solid curves) leads to a much faster decay of the norm, compared to the full complex MPS (dot-dashed curves) by about an order of magnitude.
Even for $D=200$, the SA-complex MPS has a norm of only $\numprint{0.82634974}$ at the end of the propagation whereas the full-complex MPS has  a norm of $\numprint{0.98252528}$.
(The full-complex MPS is not fully norm-conserving as two-site TDDMRG is used.) %
Thus, for the SA-complex MPS type, the eigenvector truncation of the state-averaged density matrix discussed in \autoref{sec:complex}
required for orthogonalizing to the next MPS site gives a larger error. %

While the TDDMRG with full complex MPS converges faster with bond dimension,
it also has higher computational and memory demands than TDDMRG with SA complex MPS.
To address this,
\autoref{tab:cpx-simultime} lists the average computational times per time step for the two types of complex MPS as well as total spin representations (singlet and non-singlet embedding; see \autoref{sec:singlet}). All of these timing data are extracted from shared-memory-parallelized simulations that used the same two Intel 28-Core Xeon Gold 6330 2.0 GHz CPUs.
For the same type of total spin representation---e.g.~for singlet embedding, compare the 2nd and 4th columns---the simulations with SA complex MPS are roughly a factor of $1.4$ faster than those with full complex MPS. Compared to the SA complex MPS type, the full complex MPS type leads to complex-valued renormalized operators, which increases the computational cost by up to a factor of $2$.
The actual value depends on whether the computation is memory-bandwidth-limited or not, and on
the implementation, as complex matrix multiplication can be implemented by multiplying three instead of four real-valued matrices,\rladd{\cite{3-matrix-mult-1992}}
which is exploited in \textsc{block2}.

Despite the larger computational effort of the full complex MPS compared to the SA complex MPS,
taking into account both the bond dimension convergence (\autoref{fig:complex}(a)) and the computational cost, overall, we find that the full-complex MPS offers the best ratio of accuracy over runtime.
For instance, using a full-complex MPS, the dynamics at $D=500$ can be considered converged (\autoref{fig:complex}(a), red dot-dashed curve), and the cost per time step for this simulation is
$\unit[35.2]{s}$ with singlet embedding,
whereas using a SA complex MPS, even $D=1200$ is not fully converged and  requires $\unit[145]{s}$ per time step.
\rladd{Despite the overall better performance of full complex MPS over SA complex MPS, for computational reasons, we also employ the latter type in some of our results below. Since both representations converge to the exact dynamics for a given time step in the large bond dimension limit,  the choice of MPS representation should not affect the comparison of other simulation ingredients such as the used orbitals.
To demonstrate this, some additional comparisons are shown in the SI in Section S6.}

\begin{table}
    \caption{Average wall times per TDVP propagation time step ($\Delta t = \unit[0.968]{as}$) in seconds for simulations in chloroacetylene for several bond dimensions using two different complex-valued MPS types (state averaging, SA, vs. full) and two ways of dealing with non-singlet spin states (singlet embedding, SE, or no SE).}
    \label{tab:cpx-simultime}
    \setlength{\tabcolsep}{9pt}
    \renewcommand{\arraystretch}{1.1}
    \begin{tabular}{r S[table-format=3.1] S[table-format=3.1] p{0.8mm} S[table-format=3.1] S[table-format=3.1]}
    \toprule
    \multirow[c]{2}{2em}{$D$} & \multicolumn{2}{c}{SA} & & \multicolumn{2}{c}{Full}\\
    \cline{2-3} \cline{5-6}
                              & \multicolumn{1}{c}{SE}        & \multicolumn{1}{c}{no SE}          & &      \multicolumn{1}{c}{SE} & \multicolumn{1}{c}{no SE}         \\
    \midrule
     $200$ &  10.7 &  13.5 & &  14.2 &  12.3 \\
     $500$ &  26.8 &  38.8 & &  35.2 &  50.1 \\
     $800$ &  60.8 &   110 & &  86.3 &   144 \\
    $1000$ &   101 &   173 & &   136 &   234 \\
    $1200$ &   145 &   250 & &   198 &   356 \\
    \hline
    \end{tabular}
\end{table}

\subsection{Singlet embedding (SE) \label{sec:singlet}}
\begin{figure}[!tbp]
    \includegraphics[width=\fighalf]{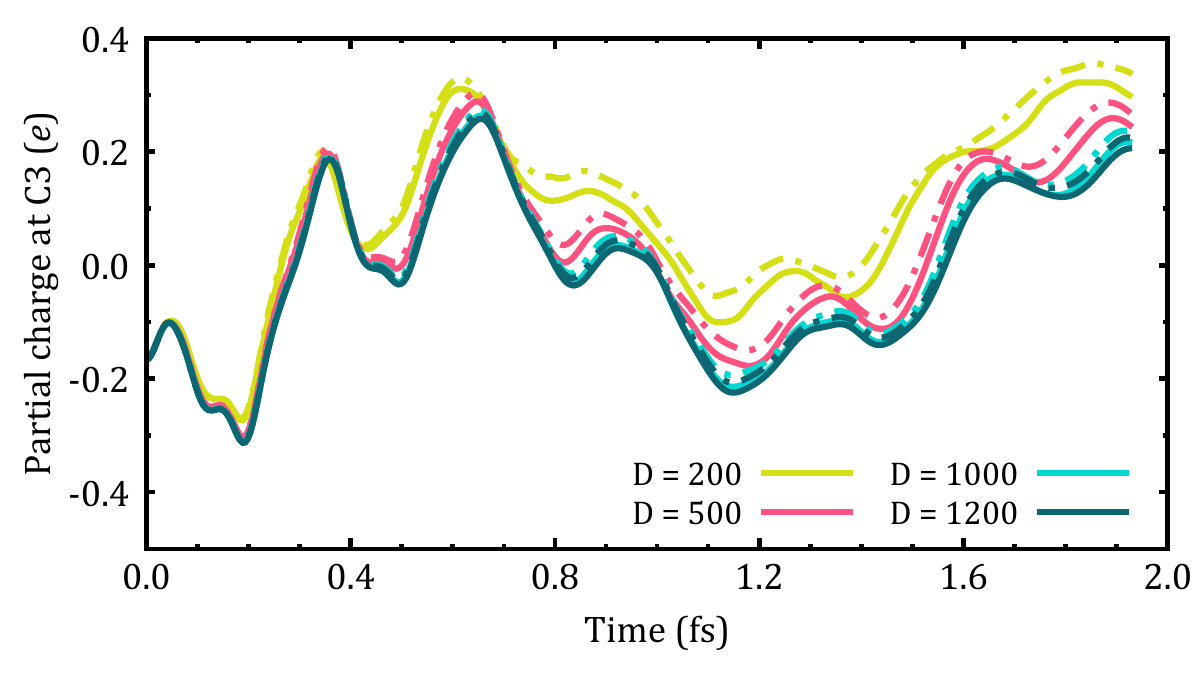}
    \caption{The effect of non-singlet spin representation in bond dimension convergence in chloroacetylene. Shown is $Q_\text{C3}(t)$ for an $SU(2)$-adapted MPS where the total spin is represented using either singlet-embedding (solid lines) or using a proper non-zero spin quantum number (dot-dashed lines). The MPS is of SA complex type.}
    \label{fig:singlet}
\end{figure}

To study the effect of singlet embedding, compared to ``proper'' non-singlet spin-adapted MPSs (see \autoref{sec:symmetry_qn}),
\autoref{fig:singlet} displays a bond dimension convergence scan for the two ways to construct open-shell spin-adapted MPSs.
Compared to the simulations using a proper spin type (dot-dashed curves), the ones with singlet-embedded MPSs (solid curves) converge slightly faster with respect to bond dimension.
This is consistent with ground-state DMRG. \cite{dmrg-spin-symm-2012}
In addition, as shown in  \autoref{tab:cpx-simultime}, compared to the proper non-singlet MPS,
TDDMRG propagation using the MPS with singlet-embedding leads to a much faster runtime.
For the converged bond dimension of $D=1200$, singlet embedding is almost twice as fast as non-singlet-embedding.

\subsection{Orbital localization: Energy or coordinate space? \label{sec:localization}}
As discussed in \autoref{sec:orbital_shape},
split localization in coordinate space typically decreases the required bond dimension for larger molecules in ground state DMRG, whereas natural orbitals (energy localization)
work best for small molecules such as diatomics.
Does this also hold for TDDMRG?
To answer this, we first benchmark localization for chloroacetylene, whose linear structure makes it well-suited for localization.
Here, we compare three different types of orbitals that are based on the previously used cationic CAS($30o,15e$):
(1) quasi-natural energy-localized orbitals, (2) split-localized orbitals, and (3) globally localized orbitals.
These orbitals are based on the hole-DM adaptation procedure described later in \autoref{sec:dyn-adapt}. In short, the quasi-natural orbitals in chloroacetylene are based on eight natural ``base'' orbitals obtained from DMRG optimization of the neutral ground state, and 22 additional quasi-natural charge orbitals that are based on an averaged hole density matrix.
The base orbitals (quasi-natural charge orbitals) also define the occupied (unoccupied) space used for the split localization.
As the molecule has linear symmetry, we localize the orbitals within each irrep sector of the $C_{2v}$ point group without jeopardizing orbital locality using the Pipek-Mezey algorithm.\cite{pm_loc-1989}
The localized orbitals are ordered by aligning them on the molecular axis whereas the quasi-natural energy-localized %
orbitals are ordered using a genetic algorithm and the default parameters in \textsc{block2}.

\begin{figure}[!tbp]
    \includegraphics[width=\fighalf]{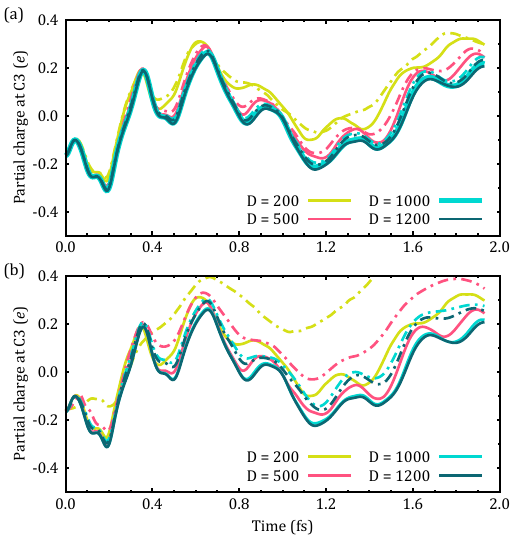}
    \caption{The effect of orbital localization in bond dimension convergence in chloroacetylene. (a) $Q_\text{C3}(t)$ using split-localized (solid lines) vs.~quasi-natural orbitals (dot-dashed lines).
    (b) $Q_\text{C3}(t)$ using split-localized (solid lines) vs. globally localized orbitals (dot-dashed lines).
    The MPS is of SA complex type.}
    \label{fig:loc}
\end{figure}

\autoref{fig:loc}(a) shows the bond dimension convergence for the partial charges at C3 when using split-localized orbitals (solid lines) and the quasi-natural orbitals (dot-dashed lines).
As expected from a larger linear molecule, \cite{dmrg_in_practice-2015, dmrg-large-site-2022}
there is a slight improvement of the convergence rate with bond dimension when split-localized orbitals are used.
For instance, the  $D=1200$ non-localized curve has a larger error than the $D=1000$ split-localized curve.
In contrast, ignoring the occupied orbital space and globally localizing all orbitals simultaneously significantly decreases the convergence rate and bond dimensions much larger than $1000$ are required for convergence, c.f.~\autoref{fig:loc}(b).
Localizing all orbitals simultaneously deteriorates ``energy locality,'' as low-energy configurations are not well-described by globally localized orbitals, thus the poor performance of those orbitals. This is in agreement with ground-state DMRG results. \cite{dmrg_in_practice-2015}

The different convergence behaviors for energy-localized, split-localized, and globally localized orbitals are also apparent in the norm conservation (not shown here). For a given bond dimension, globally localized orbitals lead to the fastest norm decay, followed by natural orbitals, and then the split-localized orbitals.

\begin{figure}[!tbp]
    \includegraphics[width=\fighalf]{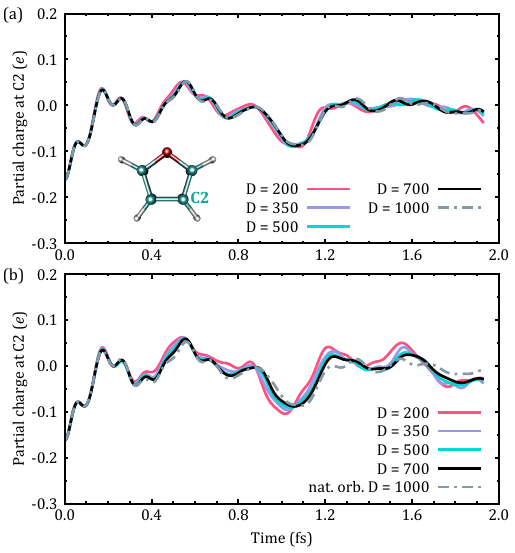}
    \caption{$Q_\text{C2}(t)$ (with C2 being the carbon atom indicated in panel a) in furan obtained using (a) MP2 natural orbitals and (b) split-localization of orbitals in (a). All results used the $C_s$ point group except for the $D=1000$ result in panel a, which used the $C_{2v}$ point group. This result is also plotted in panel b, for comparison. These simulations were performed with an active space of CAS($30o$,$25e$) and the MPS is of full complex type.}
    \label{fig:furan-loc}
\end{figure}

\begin{figure}[!tbp]
    \includegraphics[width=\fighalf]{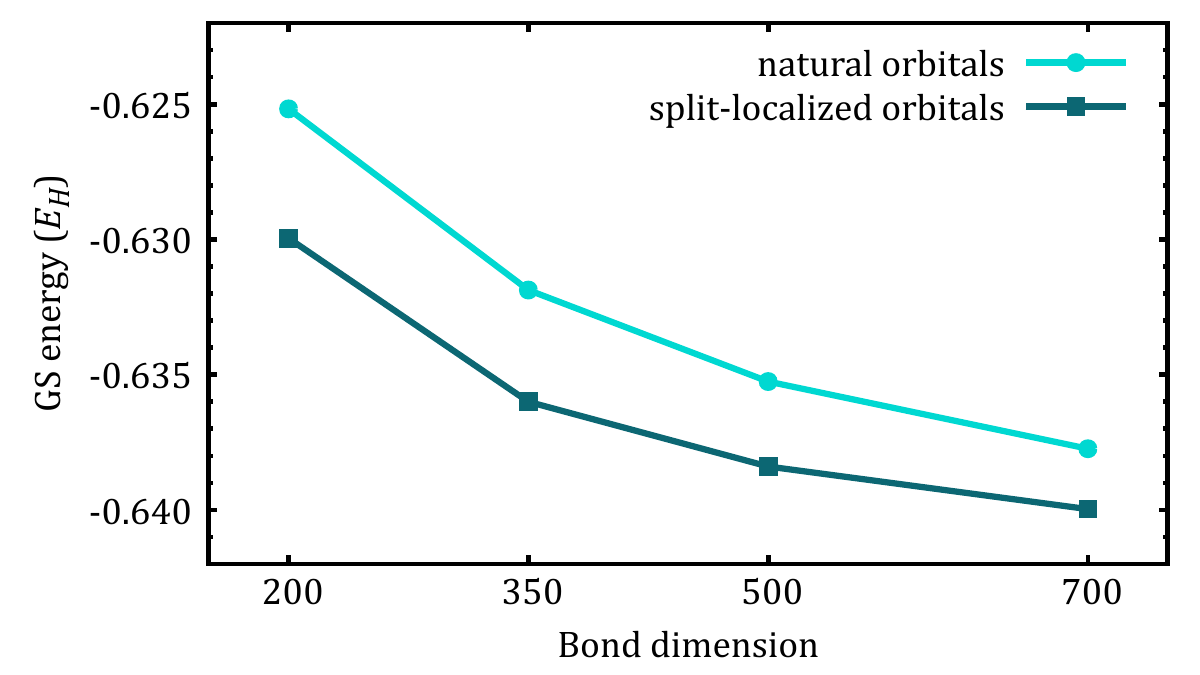}
    \caption{Ground state energy of the furan cation as function of bond dimension obtained using either MP2 natural orbitals (points) or split-localized orbitals (squares). These are the same orbitals used in  \autoref{fig:furan-loc}. The energies are shifted up by $\myunit[228]{E_H}$.}
    \label{fig:furan-erg}
\end{figure}

From the benchmark in chloroacetylene, it seems that the advantage of using split-localized orbitals to improve bond dimension convergence in ground state simulations does carry over to time-dependent dynamics.
This, however, turns out to be system-dependent, as we will show now for the case of furan. %
Here, our CAS is based on the first 30 non-core symmetric MP2 natural orbitals from the neutral molecule sorted by natural occupation.
Note that these orbitals are not the most optimal ones to describe the actual dynamics of the system, but they suffice to describe the effect of orbital localization (c.f~\autoref{sec:dyn-adapt}).
As for choloracetylene, split-localization is done within each irrep of the $C_s$ point group.
The split-localization is based on $13$ occupied and $17$ unoccupied orbitals.
We order the orbitals based on the genetic algorithm.

\autoref{fig:furan-loc} compares the bond dimension convergence
for charge migration in furan using natural orbitals (panel a), and split-localized orbitals  (panel b).
While convergence is attained very quickly in both cases,
surprisingly, the dynamics using natural orbitals (panel a) converges faster with bond dimension than that using localized orbitals (panel b).
Both orbital types lead to slightly different partial charge curves even for $D=700$.
Thus, for this dynamics, energy localization is more important than spatial localization.
This is in contrast to the ground state energy convergence of the same system (shown in \autoref{fig:furan-erg}), where split-localized orbitals clearly converge faster with respect to bond dimension.
We have found a similar behavior for dynamics in acetylene (see SI \sifig{5}), although in this case, localized orbitals also do not improve ground state energy convergence (SI \sifig{6}).
Our results on these three systems (chloroacetylene, acetylene, and furan) indicate that not all empirical findings from ground state DMRG simulations can be directly used for dynamics but rather additional benchmarks need to be performed in order to generate the most optimal setup for TDDMRG simulations.

For \rladd{furan}, natural orbitals converge faster with bond dimension
and lead to the fastest simulations if the $C_{2v}$ point group is used.
However, if  the same point group $C_s$ is used for both set of orbitals,
from the viewpoint of resource requirement, using localized orbitals here leads to a substantial resource reduction.
Compared to using natural orbitals with the $C_s$ point group, using localized orbitals  and the same point group requires almost a factor of $3$ less memory and the runtime is reduced by a factor of $2$ for $D=700$. %
This is important for molecules that are less symmetric than furan, e.g.~furfural in \autoref{sec:appl} and \sisec{5} in the SI.
Notably, this is not due to  integral sparsity, as here the natural orbitals have twice as many vanishing integrals, because the $C_{2v}$ point group is retained for natural orbitals, in contrast to the $C_s$-symmetric localized orbitals.
Instead,
the reason for the reduced resources is that, compared to natural orbitals,
the  spin- and point-group-symmetrized MPS with localized orbitals has around twice as many active blocks of the symmetrized site tensors $\matr M^{[i]}$, and each of the blocks is significantly smaller than that of the MPS with natural orbitals.
In other words, here localized orbitals lead to MPS tensors that are more block-sparse, compared to natural orbitals.
This may also explain the different convergence behavior with $D$ seen in \autoref{fig:furan-loc}, as a larger amount of block-sparsity means
that the sizes of each block need to be optimized, which is difficult and often leads to the MPS being ``trapped'' in a local minimum. \cite{dmrg_in_practice-2015}
The nature of the block sparsity is affected by both the shape and the order of the orbitals, thus an in-depth explanation of the different block-sparsities is out of the scope of this work.
However, we speculate that the MPS that is represented by natural orbitals only needs a few important orbitals to form dominant configurations through orbital rotations, which results in fewer important symmetry sectors, whereas the localized orbitals do require many different orbitals and thus also more symmetry sectors.

\section{Orbital selection \label{sec:spatial_orbital}}

Since our simulations are based on TD-CASCI and thus the orbitals are not optimized variationally at each time, selecting the most appropriate orbitals for describing the MPS for all times is crucial.
Since there is no orbital optimization, this problem is even more severe than that of finding initial guesses for CASSCF or DMRGSCF, where by now many orbital selection procedures exist, see, e.g.~\lits{orb-select1-1988, orb-select2-2001, orb-select3-2003, orb-select4-2010, natorb-for-dmrg-2013, orb-select5-avas-2017, orb-select6-2019, orb-select7-2019, orb-select8-2020, orb-select9-2023}.
Notably, the active space required for intricate dynamics simulations is typically larger than those that are adequate for ground states. \cite{hochstuhl-mctdhf-2011, mctdhf-mccurdy-2015, gasci-2016, larsson-sfi-2016}
So far, not many procedures to select the orbital space for time-dependent simulations have been considered, see
\lits{tdrasci-hochstuhl-2012, larsson-sfi-2016, levine-tdcasci-2018, cas_without_scf-2021}
for a few exceptions for some special cases.

\subsection{Dynamics-adapted orbitals \label{sec:orb_computation}}
Here, we explore two approaches to select orbitals. Both of them
are based on natural orbitals of an averaged density matrix, thus they are easy to use and systematically improvable.
Similar to some other methods, \cite{reiher-auto-orbitals-2016, orb-select7-2019, orb-select9-2023}
both are based on a preliminary TDDMRG simulation that employs all or a large enough subset of the available orbitals but uses a small bond dimension %
so that the final orbitals can be obtained quickly.
From this preliminary TDDMRG simulation, we average over the real part of the one-electron density matrix $\matrgreek \rho$ at each time step $i$  (in atomic orbital representation):
\begin{equation}
  \matrgreek \rho_\text{averaged} = \frac 1{N_\text{timesteps}} \sum_{i=1}^{N_\text{timesteps}} \Re \matrgreek{\rho}(t=i \Delta t).\label{eq:rhoav}
\end{equation}
We use the real part of $\matrgreek{\rho}(t=i \Delta t)$ to avoid dealing with complex-valued orbitals, nevertheless,
the real part still captures the main dynamics. \cite{cm-breidbach-2003}
In the first of the two approaches, dubbed here
\textit{density matrix (DM)-adapted}  orbitals, we use the natural orbitals of $\matrgreek \rho_\text{averaged}$ with the largest occupancies for the TD-CASCI simulations.
While straightforward to use, \autoref{eq:rhoav} focuses on the total average of the dynamics and thus may not easily capture subtle changes in the dynamics.
To improve this,
in the second approach, dubbed here \textit{hole-DM-adapted} orbitals, we do not diagonalize $\matrgreek \rho_\text{averaged}$ but rather diagonalize the averaged hole density matrix, $\matr h_{\text{averaged}} =
\matrgreek \rho_0 - \matrgreek \rho_{\text{averaged}}$ where $\matrgreek \rho_0$ is the density matrix of the neutral molecule (c.f.~\autoref{eq:hole_dens}).
Diagonalizing $\matr h_{\text{averaged}}$ leads to averaged
\textit{natural charge orbitals} and their corresponding eigenvalues, the \textit{hole occupations}. \cite{cm-breidbach-2003}
These orbitals thus describe the evolving hole in charge migration, and thus naturally are important ingredients of the orbital space.
As the trace of $\matr h_{\text{averaged}}$ is $1$, only a few orbitals can be selected from $\matr h_{\text{averaged}}$.
We select them based on the largest absolute value of the hole occupations, which can be negative.
The remaining orbitals, dubbed here \textit{base orbitals},
are chosen based on the natural orbitals  of the neutral ground state with occupancies close to two.
Since the natural charge orbitals are not orthogonal to the base orbitals, we symmetrically orthogonalize \cite{lowdin-non-ortho-1950, szabo1996modern-1996}
those against the base orbitals
and remove linear dependencies, if necessary. %
Typically, linear dependencies only become an issue once the used orbital space becomes very large, compared to the used atomic orbital space.
Both procedures described above allow the active space to be systematically increased.
Note that these procedures can be easily extended to other types of dynamics that are not based on ionization. In that case, instead of subtracting the density matrix of the neutral molecule to form the averaged hole density matrix, one could, e.g., subtract the density matrix of the initial state. Likewise, other established ways such as those based on localized orbitals \cite{orb-select5-avas-2017} could be used to obtain the base orbitals.
The DM-adaption procedure could be improved by using concepts other than natural orbitals, e.g., from quantum information theory. \cite{orb-select3-2003, reiher-auto-orbitals-2016, tddmrg-baiardi-2021, orb-select9-2023}
This, however, is beyond the scope of this work.
\subsection{Validation\label{sec:dyn-adapt}}

To test the two dynamics-adapted orbital approaches, here we compare two sets of simulations using the DM-adapted and hole-DM-adapted orbitals %
and vary the size of the active space both for furan and for chloroacetylene (see \autoref{sec:methods} for the main setup).
For both molecules,
we use $D=200$ for the preliminary TDDMRG simulation to obtain the averaged hole density.
Despite the small bond dimension, the simulation captures the rough qualitative behavior of the dynamics,
compared to a simulation with $D=1000$ that is used here as reference  (see SI \sifig{7}).

For the hole-DM-adapted orbitals of chloroacetylene,
we use eight base orbitals from the natural orbitals of an RDM obtained from a frozen-core FCI DMRG neutral ground state calculation with $D=200$. %
We use split localization (c.f.~\autoref{sec:localization})
to optimize the orbital shape. Split localization of the hole-DM-adapted orbitals is done using the base orbitals as occupied space and the remaining orbitals as valence space.
For the DM-adapted orbitals, split localization is based on
the quasi-natural orbitals of the averaged RDM in \autoref{eq:rhoav} using a natural occupation threshold of $0.5$.

\begin{figure}[!tbp]
    \includegraphics[width=\fighalf]{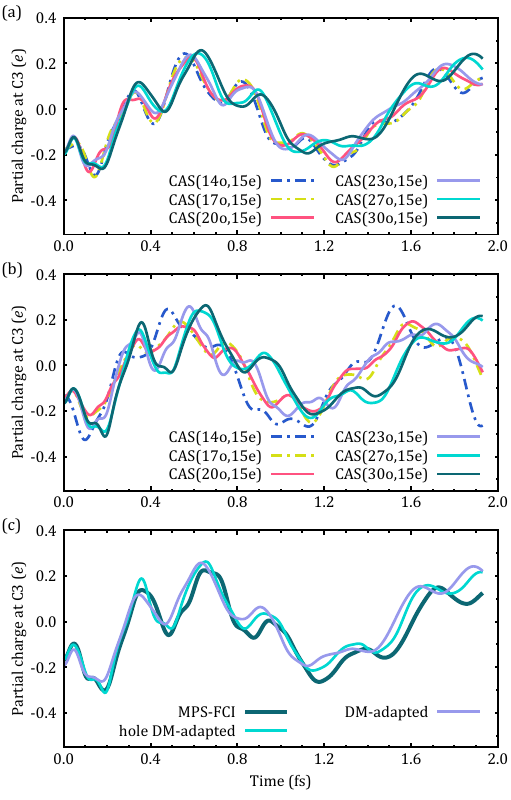}
    \caption{$Q_\text{C3}(t)$ in chloroacetylene obtained using several active spaces composed of (a) DM-adapted orbitals and (b) hole-DM-adapted orbitals. The simulations with the largest CAS($30o,15e$) are compared to MPS-FCI result,  a CAS($41o,15e$), in panel (c). The bond dimension is $1000$ and the MPS is of SA complex type.}
    \label{fig:cas}
\end{figure}

\autoref{fig:cas} displays the results for chloroacetylene. %
The DM-adapted orbitals (\autoref{fig:cas} a) capture many suboscillations of the charge even if the active spaces are relatively small, such as only 14 orbitals. The overall shapes of the curves for all CASs shown are very similar, up to some horizontal shifts.
On the contrary, the hole-DM-adapted orbitals (\autoref{fig:cas} b) produce dynamics that varies greatly as the active space size is increased. Convergence is attained only when reaching $\sim$27 orbitals out of 41 in total.
Importantly, however, a comparison of the partial charges of the simulations using either DM-adapted orbitals or hole-DM-adapted orbitals with the FCI TDDMRG reference (dubbed MPS-FCI) (\autoref{fig:cas}c) reveals that when including 30 orbitals in the space, the hole-DM-adapted orbitals lead to dynamics that is actually closer to MPS-FCI than the DM-adapted orbitals, despite the former's poor performance when the active space is small.

Next to the two DM-adapted procedures, we have tested three other ways of selecting orbitals for the active space: (1) DMRGSCF orbitals of the neutral molecule, (2) DMRGSCF orbitals of the cation, and (3) state-averaged DMRGSCF orbitals of the cation.
In all three procedures, the initial DMRGSCF guess is based on split-localized MP2 natural orbitals.
These simulations are shown in the SI (\sifig{8}) and confirm that, for chloroacetylene, the CAS created by hole-DM-adapted orbitals leads to the fastest convergence. %

\begin{figure}[!tbp]
    \includegraphics[width=\fighalf]{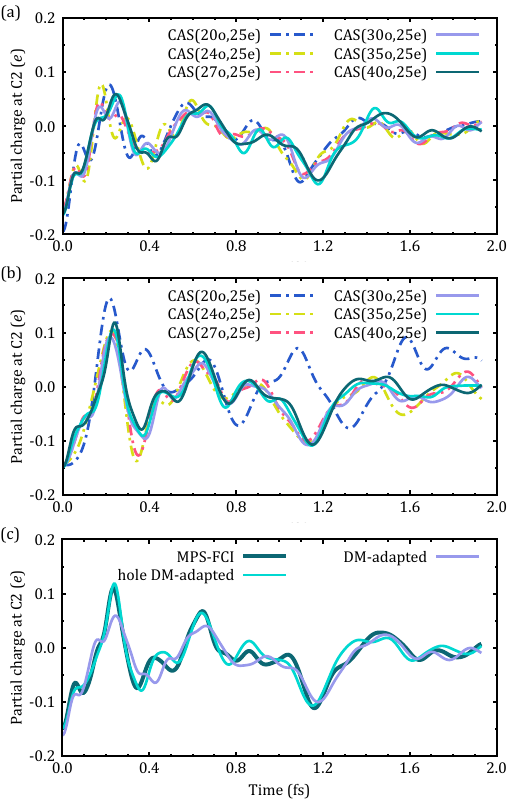}
    \caption{$Q_\text{C2}(t)$ in furan obtained using several active spaces composed of (a) DM-adapted orbitals and (b) hole-DM-adapted orbitals. (c) A comparison of the CAS($40o,25e$) results from panel (a) and (b) with the MPS-FCI result for this molecule, whose active space is CAS($48o,25e$).
    See  \autoref{fig:furan-loc} for details on $Q_\text{C2}(t)$.
    The bond dimension is $500$ and the MPS is of full complex type.}
    \label{fig:furan-cas}
\end{figure}

The results for the same type of comparison for furan are shown in \autoref{fig:furan-cas}.
For the hole-DM-adapted orbitals, we use 13 base orbitals computed in the same way as for chloroacetylene, except for the DMRG neutral ground state computation, which uses $D=1000$. %
Since for furan, energy-localized orbitals lead to a faster convergence with respect to bond dimension (see \autoref{sec:localization}), we use the quasi-natural orbitals from the DM adaption procedures directly and order them using a genetic algorithm.
Here, the  DM-adapted orbitals (\autoref{fig:furan-cas} a) converge slower than  the hole-DM-adapted orbitals (\autoref{fig:furan-cas} b)
with respect to active space size, which is in contrast to our observation in choloracetylene.
However, as in chloroacetylene, the hole-DM-adapted orbitals yield a better agreement with MPS-FCI than the DM-adapted orbitals, suggesting that a faster convergence with active space may be realized by using hole-DM-adapted orbitals.

Overall, thus we find that hole-DM-adapted orbitals are preferred over DM-adapted orbitals, which is in agreement with our discussion in  \autoref{sec:orb_computation}.
Based on the results for chloroacetylene, which  do not display smooth convergence, the hole-DM adaptation is not a full black-box method. As the orbitals are not time-dependent, it is expected that erratic results can be obtained if not all important orbitals are included in the space.
Nevertheless, e.g.~the use of concepts from quantum information theory, as mentioned in \autoref{sec:orb_computation}, would allow for additional improvements.

\section{Applications---dynamics resulting from $\sigma$ and $\pi$ orbitals ionization in furfural}
\label{sec:appl}

In this section, we will apply the knowledge about the numerical behavior of TDDMRG learned from \autoref{sec:benchmark} to furfural to observe charge migration from the formyl group into the five-membered ring. %
We use the 6-31G basis set, giving $71$ basis functions for furfural with seven core orbitals kept frozen. The used geometry is given in \sisec{1} of the SI.

We study two scenarios, where either a $\pi$-bonding or a $\sigma$-bonding IBO of the formyl group is ionized. We will call these two scenarios \textit{$\pi$ dynamics}, and \textit{$\sigma$ dynamics}, respectively.
The $\pi$ ($\sigma$) dynamics corresponds to outer (inner) valence ionization. %
Based on the previous findings, we use hole-DM-adapted orbitals
and found a CAS($40o,35e$) and $D=700$ for a complex MPS to be sufficiently converged for both scenarios and for the observables of interest (see below). The time steps used for the $\pi$ and $\sigma$ dynamics are $\unit[0.968]{as}$ and $\unit[0.484]{as}$, respectively.
The preliminary TDDMRG simulations to obtain the two sets of orbitals
use a bond dimension of $200$. %
A convergence analysis for the dynamics is available in \sisec{5.1} of the SI, in particular, see  Figs.~S9-S11.

\begin{figure}[!tbp]
    \includegraphics[width=\fighalf]{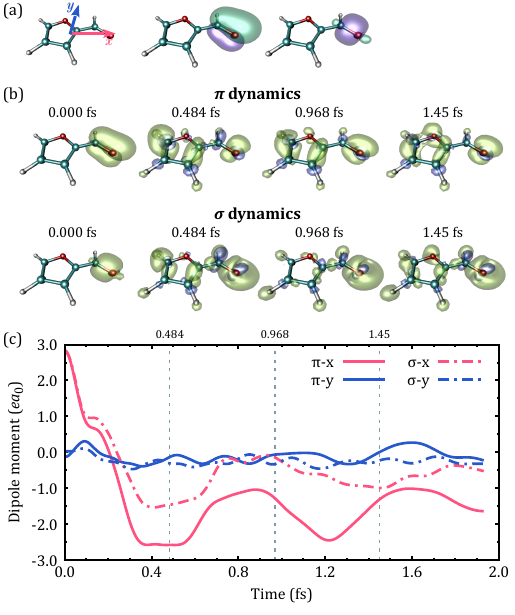}
    \caption{Charge migration in furfural based on ionization from either a $\pi$ or a $\sigma$ orbital.
    (a) Left panel: Geometry with the reference frame used for calculating the dipole moments (principal axes of inertia).
    Right panels:
    Isosurface of the orbitals from where the electron is ionized at  $|\psi(\mathbf r)|=0.06$.
    (b) Hole density isosurfaces evaluated at $h(\mathbf{r},t)=\pm0.006$. The green (violet) surfaces show positive (negative) values.
    (c) Total dipole moments in $x$ and $y$ as function of time for the two scenarios. Solid (dot-dashed) lines show ionization from the $\pi$ ($\sigma$) orbital.
    Due to symmetry, the  $z$-component of the dipole vanishes.
    The vertical lines mark the times at which the hole densities in panel (b) are evaluated.
    }
    \label{fig:furfural-dipole}
\end{figure}

Here, we are interested in the general dynamics and not in particular benchmarks. Thus, we study two observables different to those in \autoref{sec:benchmark}, namely the time-dependent dipole moments and the hole density.
The dipole moments are calculated along the principal axes of inertia. %
The observables, the used coordinates, and the initial states are shown in  \autoref{fig:furfural-dipole}, next to the two IBOs used for the ionization scenarios (panel a).
Despite the different shapes and symmetries of the initial hole densities (panel b), at first sight the initial dynamics of the resulting charge migration %
is surprisingly similar.
However, a closer look reveals more subtle differences.
In particular, the $\sigma$ dynamics leads to a reduced migration with larger parts of the density staying in the formyl group, whereas for the $\pi$ dynamics
a larger portion of the initial hole migrates into the ring.

This is confirmed by inspecting the dipole moments (panel c).
In the $x$ direction, the dipole changes from positive  to negative values (the initial migration into the ring) and then oscillates around a negative mean value.
Like the apparent similarities of the time-dependent hole densities, the oscillation periods of the $x$ dipole components of the $\sigma$ and $\pi$ dynamics during the first $\unit[0.8]{fs}$ are almost equal, even though the energy of the initial state of the $\sigma$ dynamics is $\unit[18]{eV}$ higher than that of the $\pi$ dynamics.
However,
both $\sigma$ and $\pi$ dynamics show these oscillations with a noticeably smaller initial shift of the dipole moment for the $\sigma$ dynamics from $\myunit[2.80]{ea_0}$ to $\myunit[-1.53]{ea_0}$, compared to  $\myunit[2.84]{ea_0}$ to $\myunit[-2.57]{ea_0}$ for the $\pi$ dynamics.
At later times, each of the two dynamics shows a  different behavior, e.g., at $\unit[1.45]{fs}$ the dipole in $x$ direction of the $\sigma$ dynamics is around a minimum  whereas that of the $\pi$ dynamics approaches a maximum.
As the direction of the initial charge migration in the molecule can only occur from the formyl group into the ring, the $y$ component of the dipole is less affected by the dynamics and approximately is zero throughout. %

The observed effects might be explained by a similar, albeit reduced overlap between the hole orbital and the
Restricted Open-Shell Hartee-Fock (ROHF) cationic orbitals in the ring: %
The initial $\pi$ hole is mostly similar to one $\pi$ ROHF
orbital of the furfural cation, whereas the initial $\sigma$ hole has large overlaps to many ROHF orbitals, both in the inner valence and outer valence regions. The initial $\sigma$ hole state thus has not only a much higher energy but also a much larger energy bandwidth.
The dominant configurations of the $\sigma$ dynamics then relax to configurations that are also dominant in the  $\pi$ dynamics after a short time interval following the start of the evolution.
Note that orbitals of different symmetry are coupled through excited configurations,
thus, e.g., $\pi\pi^\ast$ excitations can appear during charge migrations of initial $\sigma$ holes.
A similar effect has been observed previously for a simpler scenario, see, e.g., \lit{cm-breidbach-2003}.

In agreement with the above explanation, an analysis of the natural charge orbitals (see SI Figs.~S12-S14)
at $t=\unit[0.484]{fs}$ reveals that $\pi$ orbitals are active components not only in the $\pi$ dynamics but also in the $\sigma$ dynamics.
As the dynamics advance, the charge orbitals become less similar: at $\unit[0.968]{fs}$ both dynamics have a dominant $\pi$-type charge orbital localized at the formyl group, whereas at $\unit[1.45]{fs}$
the same type of  orbital remains important only for the $\sigma$ dynamics.
This decreasing similarity between the two dynamics as time advances is in agreement with the hole densities and dipoles from \autoref{fig:furfural-dipole}.
\section{Conclusions}%
\label{sec:conclusions}
Time-dependent extensions of the density matrix renormalization group (TDDMRG) are promising candidates for efficient simulations of complicated molecular electron real-time dynamics.
However, so far only a few TDDMRG applications exist and, due to a workflow that is more complicated than ground-state optimizations, so far there are only a few guidelines on choosing optimal simulation parameters.
Here, we have aimed for establishing a robust and general TDDMRG simulation framework.
Through simulations of charge migration in different molecules,
we have performed a series of extensive studies of the properties of various simulation parameters in TDDMRG, and investigated ways to determine compact orbital spaces for the electron dynamics.

We have found that the time-dependent variational-principle-based projector splitting integrator converges
faster and more smoothly than the time-step targeting method with both bond dimension and time step.
Using a fully complex-valued   matrix product state (MPS) is more elaborate than a state-averaged-based complex-valued representation, but the fully complex-valued MPS  converges faster with bond dimension and converges the norm of the state well, even if the two-site version of the DMRG is used.

One of the questions we have aimed to address was whether common wisdom from ground-state DMRG carries over to TDDMRG. Indeed, we have found that some ground-state-based findings do carry over to TDDMRG, namely orbital ordering, and the use of singlet embedding for non-singlet states.
However, surprisingly we found that %
split-localized orbitals does not necessarily speed up bond dimension convergence for larger molecules, compared to natural orbitals. This is in contrast to the convergence behavior of ground state DMRG.

Next to the benchmark, we have tested several procedures to obtain a good set of active orbitals.
This is more difficult than in time-independent DMRG as the orbitals need to describe the dynamics well for all propagation times.
We found that orbital spaces consisting of a combination of occupied ground-state natural orbitals and orbitals that diagonalize a
reduced density matrix (RDM) consisting of the difference between an averaged time-dependent RDM and a ground-state RDM lead to reasonably fast convergence.
This procedure offers a semi-automated way to converge the active space.

Lastly, we have made use of our findings in the simulations of charge migration in furfural, where we compared the dynamics of $\pi$- and $\sigma$-hole-initiated charge migrations. We found a rapid conversion of the initial $\sigma$ hole to $\pi$ hole states.
These simulations employ $35$ electrons and $40$ active orbitals, and the dynamics displays nontrivial behavior, rendering them significantly  more challenging than previous molecular electronic TDDMRG applications.
Given the efficiency of TDDMRG to simulate multi-reference situations, we believe that it will be more widely used to study more intricate dynamics, such as when an external field is present, or to interpret experimental observations.

\section*{Associated Content}
\rladd{Supporting Information available. Details on the used geometries.
Bond dimension convergence for TDVP and TST,   benchmarks of the SIL tolerance in TDVP and number of sub-sweeps in TST, and orbital localization effects in acetylene. 
Additional data on the choice of orbitals obtained using several different methods not shown in the main text. Natural charge orbital analysis for the $\pi$ and $\sigma$ dynamics in furfural. Additional convergence benchmarks for the propagation type, orbital localization effects, and the use of dynamics-adapted orbitals using SA complex representation.}

\if\USEACHEMSO1
\begin{acknowledgement}
\else
\acknowledgements
\fi
We thank Huanchen Zhai for helpful discussions.
This work was supported by the American Chemical Society Petroleum Research Fund
via grant no.~67511-DNI6.
This work was partially supported by 
\rladd{the US
National Science Foundation (NSF) via grant no.~CHE-2312005,}
University of California Merced start-up funding, %
and through
computational time on the Pinnacles and Merced clusters at University of California Merced (supported by NSF OAC-2019144 and ACI-1429783).
\if\USEACHEMSO1
\end{acknowledgement}
\fi

\providecommand{\noopsort}[1]{}\providecommand{\singleletter}[1]{#1}%

\if\USEACHEMSO1
\clearpage
{\centering
\includegraphics{toc_graphic.pdf}}
\fi

\end{document}